\documentclass[11pt,twoside]{article}  % Leave intact
\usepackage{adassconf}

\begin{document}   % Leave intact

%-----------------------------------------------------------------------
%			    Paper ID Code
%-----------------------------------------------------------------------
% Enter the proper paper identification code.  The ID code for your
% paper is the session number associated with your presentation as
% published in the official conference proceedings.  You can           
% find this number locating your abstract in the printed proceedings
% that you received at the meeting or on-line at the conference web
% site; the ID code is the letter/number sequence proceeding the title 
% of your presentation. 
%
% This will not appear in your paper; however, it allows different
% papers in the proceedings to cross-reference each other.  Note that
% you should only have one \paperID, and it should not include a
% trailing period.
%
% EXAMPLE: \paperID{O4-1}
% EXAMPLE: \paperID{P7-7}
%

\paperID{O10.2}

%-----------------------------------------------------------------------
%		            Paper Title 
%-----------------------------------------------------------------------
% Enter the title of the paper.
%
% EXAMPLE: \title{A Breakthrough in Astronomical Software Development}
% 
% If your title is so long as to fill the page header when you print it,
% then please supply a short form as a \titlemark.
%
% EXAMPLE: 
%  \title{Rapid Development for Distributed Computing, with Implications
%         for the Virtual Observatory}
%  \titlemark{Rapid Development for Distributed Computing}
%

\title{Constructing the Subaru Advanced Data and Analysis Service on VO}
%\titlemark{ }

%-----------------------------------------------------------------------
%		          Authors of Paper
%-----------------------------------------------------------------------
% Enter the authors followed by their affiliations.  The \author and
% \affil commands may appear multiple times as necessary (see example
% below).  List each author by giving the first name or initials first
% followed by the last name.  Authors with the same affiliations
% should grouped together. 
%
% EXAMPLE: \author{Raymond Plante, Doug Roberts, 
%                  R.\ M.\ Crutcher\altaffilmark{1}}
%          \affil{National Center for Supercomputing Applications, 
%                 University of Illinois Urbana-Champaign, Urbana, IL
%                 61801}
%          \author{Tom Troland}
%          \affil{University of Kentucky}
%
%          \altaffiltext{1}{Astronomy Department, UIUC}
%
% In this example, the first three authors, "Plante", "Roberts", and
% "Crutcher" are affiliated with "NCSA".  "Crutcher" has an alternate 
% affiliation with the "Astronomy Department".  The fourth author,
% "Troland", is affiliated with "University of Kentucky"

\author{Yuji Shirasaki, Masahiro Tanaka, Satoshi Honda, Satoshi Kawanomoto,
        Masatoshi Ohishi, Yoshihiko Mizumoto}
\affil{National Astronomical Observatory of Japan,
       2-21-1 Osawa, Mitaka Tokyo, 181-8588 Japan}

\author{Naoki Yasuda}
\affil{University of Tokyo, 5-1-5 Kashiwa-no-Ha, Kashiwa Chiba, 277-8582 Japan}

\author{Yoshifumi Masunaga}
\affil{Ochanomizu Univerisity, 2-1-1 Otsuka Bunkyo-ku, Tokyo, 112-8610 Japan}

\author{Yasuhide Ishihara, Jumpei Tsutsumi}
\affil{Fujitsu Ltd., 4-1-1 Kamikodanaka Nakahara-ku, Kawasaki, 211-8588}

\author{Hiroyuki Nakamoto, Yuusuke Kobayashi, Michito Sakamoto}
\affil{Systems Engineering Consultants Co. Ltd., 22-4 Sakuraoka-cho Shibuya-ku,
       Tokyo, 150-0031}

%-----------------------------------------------------------------------
%			 Contact Information
%-----------------------------------------------------------------------
% This information will not appear in the paper but will be used by
% the editors in case you need to be contacted concerning your
% submission.  Enter your name as the contact along with your email
% address.
% 
% EXAMPLE:  \contact{Dennis Crabtree}
%           \email{crabtree@cfht.hawaii.edu}
%

\contact{Yuji Shirasaki}
\email{yuji.shirasaki@nao.ac.jp}

%-----------------------------------------------------------------------
%		      Author Index Specification
%-----------------------------------------------------------------------
% Specify how each author name should appear in the author index.  The 
% \paindex{ } should be used to indicate the primary author, and the
% \aindex for all other co-authors.  You MUST use the following
% syntax: 
%
% SYNTAX:  \aindex{Lastname, F. M.}
% 
% where F is the first initial and M is the second initial (if
% used).  This guarantees that authors that appear in multiple papers
% will appear only once in the author index.  
%
% EXAMPLE: \paindex{Crabtree, D.}
%          \aindex{Manset, N.}        
%          \aindex{Veillet, C.}        
%
% NOTE: this information is also used to build the author list that
% appears in the table of contents.  Authors will be listed in the order
% of the \paindex and \aindex commmands.
%

\paindex{Shirasaki, Y.}
\aindex{Tanaka, M.}
\aindex{Honda, S.}
\aindex{Kawanomoto, S.}
\aindex{Ohishi, M.}
\aindex{Mizumoto, Y.}
\aindex{Yasuda, N.}
\aindex{Masunaga, Y.}
\aindex{Ishihara, Y.}
\aindex{Tsutsumi, J.}
\aindex{Nakamoto, H.}
\aindex{Kobayashi, Y.}
\aindex{Sakamoto, M.}

%-----------------------------------------------------------------------
%		      Author list for page header	
%-----------------------------------------------------------------------
% Please supply a list of author last names for the page header. in
% one of these formats:
%
% EXAMPLES:
% \authormark{Lastname}
% \authormark{Lastname1 \& Lastname2}
% \authormark{Lastname1, Lastname2, ... \& LastnameN}
% \authormark{Lastname et al.}
%
% Use the "et al." form in the case of seven or more authors, or if
% the preferred form is too long to fit in the header.

\authormark{Shirasaki et al.}

%-----------------------------------------------------------------------
%			Subject Index keywords
%-----------------------------------------------------------------------
% Enter a comma separated list of up to 6 keywords describing your
% paper.  These will NOT be printed as part of your paper; however,
% they will be used to generate the subject index for the proceedings.
% There is no standard list; however, you can consult the indices
% for past proceedings (http://adass.org/adass/proceedings/).
%
% EXAMPLE:  \keywords{visualization, astronomy: radio, parallel
%                     computing, AIPS++, Galactic Center}
%
% In this example, the author noticed that "radio astronomy" appeared
% in the ADASS VII Index as "astronomy" being the major keyword and
% "radio" as the minor keyword.  The colon is used to introduce another
% level into the index.


%-----------------------------------------------------------------------
%			       Abstract
%-----------------------------------------------------------------------
% Type abstract in the space below.  Consult the User Guide and Latex
% Information file for a list of supported macros (e.g. for typesetting 
% special symbols). Do not leave a blank line between \begin{abstract} 
% and the start of your text.

\begin{abstract}          % Leave intact

We present our activity on making the Subaru Data Archive accessible
through the Japanese Virtual Observatory (JVO) system.
There are a lot of demand to use the archived Subaru data from various
fields of astronomers.
To be used by those who are not familiar with the way to reduce the
Subaru data, the data reduction should be made before providing for
them, or at least it should be easily done without precise knowledge 
about instrument's characteristic and environment where data are taken.
For those purposes, data quality assessment system NAQATA is developed,
which is presented in this meeting by Nakata et al. (2006), and the
science-ready image data are provided for some of the SuprimeCam data at
SMOKA data service which is presented by Enoki et al (2006).
JVO portal will provide the way to access the reduce data, and also
provides the way to reduce from raw data with very few efforts through 
the user-friendly web browser I/F.
To provide such a CPU-intensive service, we have developed a GRID
computing system.
The architecture of this Subaru Data and Analysis system are discussed.

\end{abstract}

%-----------------------------------------------------------------------
%			      Main Body
%-----------------------------------------------------------------------
% Place the text for the main body of the paper here.  You should use
% the \section command to label the various sections; use of
% \subsection is optional.  Significant words in section titles should
% be capitalized.  Sections and subsections will be numbered
% automatically. 
%
% EXAMPLE:  \section{Introduction}
%           ...
%           \subsection{Our View of the World}
%           ...
%           \section{A New Approach}
%
% It is recommended that you look at the sample papers, sample1.tex
% and sample2.tex, for examples for formatting references, footnotes,
% figures, equations, html links, lists, and other special features.  

\section{Introduction}

Thanks to the progress of telescope technology and the detection 
technique of recent years, it is expected that we will meet with 
a situation where a large scale of high-quality data is continuously 
generated by such as Subaru Telescope, Sloan Digital Sky Survey an so on.
The way of traditional analysis, however, appears to be insufficient 
for using the large amount of data effectively and efficiently and
getting the maximum scientific results.
Although many people recognize the importance of research that uses
the multi-wavelength data, such research actually needs considerable 
effort.
One reason is that, for each data set, one needs to learn how to reduce
and analyze the data, and even needs to know where the analysis tools
are available.
To overcome such situation and maximize the scientific return from a
big project like Subaru and ALMA, it is important to construct an 
environment where user can access to the science-ready data with 
very few effort.
National Astronomical Observatory of Japan (NAOJ) started its VO project
(Japanese Virtual Observatory -- JVO) in 2002.
The objectives of the JVO project are to provide a seamless access to 
the distributed data service in the VO, and to provide user-friendly 
analysis environment.
This paper describes our recent progress on the second objective.

\section{Current Status of the Subaru Data Archive}

\htmladdnormallinkfoot{Subaru}{http://subarutelescope.org/},
is an optical-infrared 8.2~m telescope operated by National
Astronomical Observatory of Japan (NAOJ) at Mt. Mauna Kea Hawai.
Subaru has seven open use instruments:
CIAO, 
COMICS, 
FOCUS, 
IRCS,
SuprimeCam, 
HDS and
MOIRCS.
CISCO is no longer available for open-use.
Using these instruments, observation can be made for wavelengths from
optical (300 nm) to infrared (20 $\mu$m) with spectrum resolution up to
$10^{5}$ (HDS).

\begin{figure}[b]
\epsscale{1.0}
\plotone{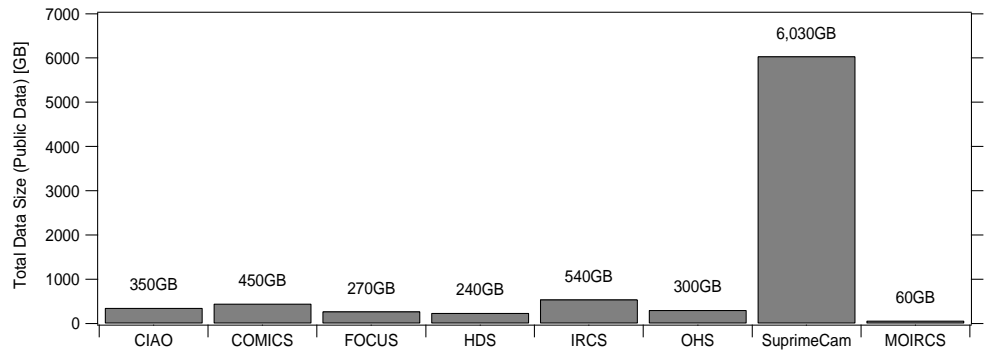}
\caption{Total amount of public data for each instrument} 
\label{fig:totalAmount}
\end{figure}
The total amount of public data for each instrument
is shown in Figure~\ref{fig:totalAmount}.
As of October 2006, 8 TB of data is archived in the public area.
More than 70\% of the data are from the SuprimeCam, which is a mosaic 
of ten 2048 $\times$  4096 CCDs and covers a 34' x 27' field of view with
a pixel scale of 0.20''.
More than 90\% of all the data requests are for the SuprimeCam, so our
current priority issue is how to improve the usability of the SuprimeCam
data.

\begin{figure}
\epsscale{1.0}
\plotone{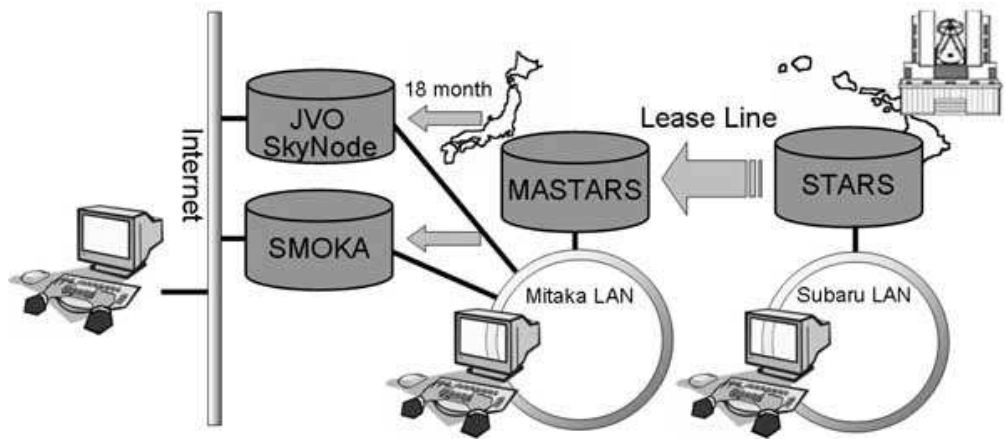}
\caption{Data flow of Subaru data} 
\label{fig:subaruArchive}
\end{figure}

Figure~\ref{fig:subaruArchive} shows the flow of data taken by the Subaru.
The data are registered in the Subaru Telescope Archive System (STARTS) 
as soon as the data are acquired by the instruments, so an observer retrieves
his data from STARS during and/or after the observation.
The data of STARS is mirrored to the Mitaka Advanced STARS (MASTARS), so
the observer can retrieve the data also from the MASTARS.
STARTS and MASTARS are not public data archive.
To use the system, you need to get an account on the Subaru computing system
for STARS or an account on the Mitaka computing system for MASTARS.
The data that passed 18 months of a proprietary period becomes publicly
available through the 
\htmladdnormallinkfoot{SMOKA}{http://smoka.nao.ac.jp/},
and 
\htmladdnormallinkfoot{JVO}{http://jvo.nao.ac.jp/}
system.
The SMOKA system provides various query modes for the Subaru archive, and it
is described in Enoki et al. (2006).
The JVO system provides a VO standard access interface to the Subaru archive.
Currently only the data of SuprimeCam is available from the JVO.

The amount of data, especially of SuprimeCam, is very large, so it is important
to provide a way to analyze the data without moving the data to a remote
users' machine.
One of the ways to do so is to login to the Subaru or Mitaka computing system 
and analyze the data on the machine.
It is, however, not practical to use the visualization tool from a remote 
machine, especially when accessing through a slow network.
Another solution is to provide a web service, through which one can access to the
data analysis software and visualize the data in a compact graphical format such
as GIF, JPEG and PNG.
Recently a lot of open source framework are available for making such a service,
and it has been realized that interactivity of the web based service can be 
improved by using Ajax technique as demonstrated by the google map service.

Another important point is to provide the data in a form that detector and 
environment dependencies are removed.
Those dependencies are usually not known by an archive user, so it should be
properly reduced by a data provider.
The quality of the data reduction is improved as the experience is
accumulated and the reduction software is also evolves continuously,  so
it is adequate to reduce the data on demand with the most developed
algorithm.
It is also important to provide the way to reduce the data with older
version software for assuring data reproducibility to enable later same
analysis.
By providing the data in such a manner, a data provider can control the quality
of data by putting a tag representing a version of reduction software in
the FITS header.
The reduction process is hidden from a user, and the user does not need
to take care about most of part of data reduction.
So, we have decided to provide such a service on the Japanese Virtual Observatory
(JVO) web portal.

\section{Grid Computing System}

\begin{figure}
\epsscale{0.9}
\plotone{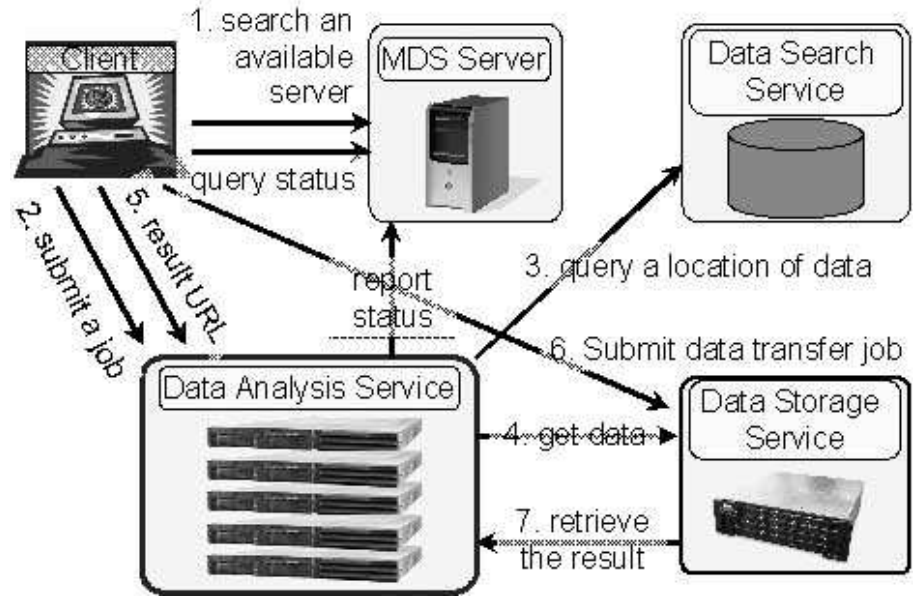}
\caption{Subaru data reduction pipeline architecture} 
\label{fig:subaruPipeline}
\end{figure}

The data reduction processes are concentrated on the JVO servers, so
computing resources need to be integrated in a scalable manner to the JVO.
We have developed a Web service based grid computing system.
This system is composed of four services:
Monitoring and Discovering Service (MDS), 
Data Analysis Service (DAS),
Data Search Service (DSS) and
Storage Service (SRS).
Figure~\ref{fig:subaruPipeline} shows an example of SuprimeCam response (flat 
frame) calculation system, which is showing how each service interacts each 
other.
The MDS manages the status of each DAS, and schedules the job 
submission requested by a GRID client.
Each DAS server periodically report its status, such as load average,
number of running job and job status, and the status is stored on an
MDS database.
A GRID client queries to the MDS to ask which server is free for a job
submission.
The MDS returns a service endpoint URL that is appropriate for the job
submission.
The client submit the job, and wait it to finish.
While waiting, the client periodically polling to the MDS for querying the
job status.
The DAS sends a query to the DSS to get Subaru RAW data, the DSS returns
the URLs for the data, the DAS retrieves the data and starts analysis.
When a job status is changed to ``finish'', the client queries the DAS
for an URL to retrieve the result.
The URL is passed to the SRS, the result is stored on the storage of 
the SRS, and the metadata of the result is registered on the DSS.

The interactions between each service and client are made by exchanged
a SOAP message.
The interface of each service is defined in the Web Service Description 
Language (WSDL).
Several examples of the defined interfaces are described in Java
interface as follows:
\begin{verbatim}
MDS Interface
   void reportStatus(String hostId, double load, int njob)
   ServiceInfo resolveService(String serviceId)
   void reportJobStatus(String hostId, int jobId, String status)
   String getJobStatus(String hostId, int jobId)
   ...

DAS Interface
   int submitJob(String command, String argv)
   String getResultURL(int jobId)
   String query(int jobId)
   String finalize(int jobId)
   ...

SRS Interface
   int copyAsync(String src, String dest)
   void copy(String src, String dest)
   void finalize(int jobId)
   ...

DSS Interface
   VOData performQuery(Select select)
   ...
\end{verbatim}
The \verb|reportStatus| interface is used by a DAS server to report its
load average and the number of submitted jobs. 
The \verb|resolveService| interface is used by a GRID client to decide
which server to submit a job.
The returned \verb|ServiceInfo| contains an endpoint URL.
The \verb|submitJob| interface is used to submit a job to the DAS
server, which returns a job ID and it is used for polling the job status.
The \verb|getResultURL| interface returns an access URL for retrieving
the result.
The \verb|copyAsync| and \verb|copy| interface are used to transfer data
between two servers or just to copy the data inside the same server.
The former interface is used for data transfer that is expected to take
long time.
The \verb|performQuery| interface is used for querying the data by
Astronomical Data Query Language 
(\htmladdnormallinkfoot{ADQL}{http://www.ivoa.net/twiki/bin/view/IVOA/IvoaVOQL}),
and return the result in 
\htmladdnormallinkfoot{VOTable}{http://www.ivoa.net/twiki/bin/view/IVOA/IvoaVOTable}
format.
%
% We have a plan to adapt the NAREGI GRID middledware to this system.
%

\begin{figure}[t]
\epsscale{1.0}
\plotone{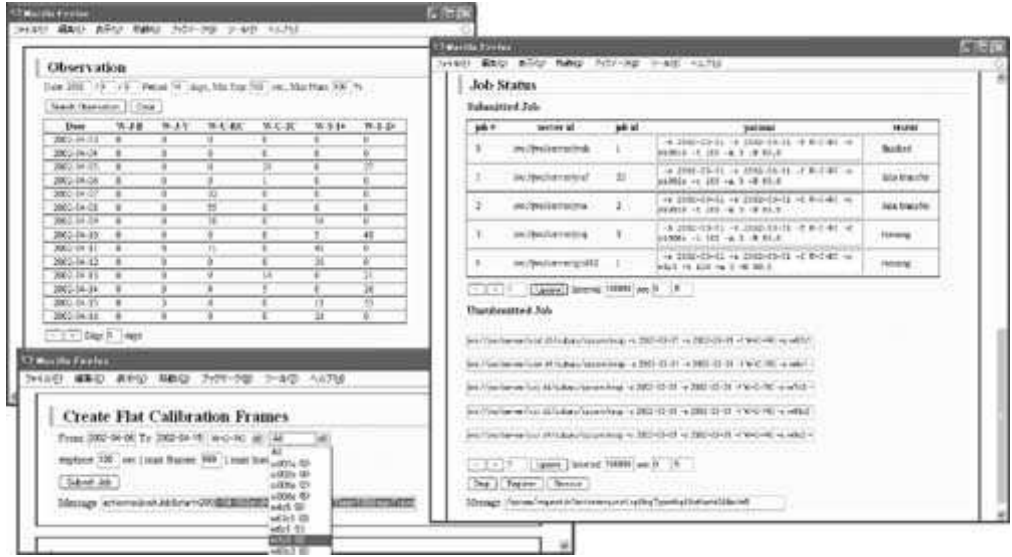}
\caption{Subaru response calculator GUI} 
\label{fig:subaruFlat}
\end{figure}
A web form based GUI for calculating a response frame of SuprimeCam CCD
is made for easy job submission as shown in Figure~\ref{fig:subaruFlat}. 
\begin{figure}
\epsscale{1.0}
\plotone{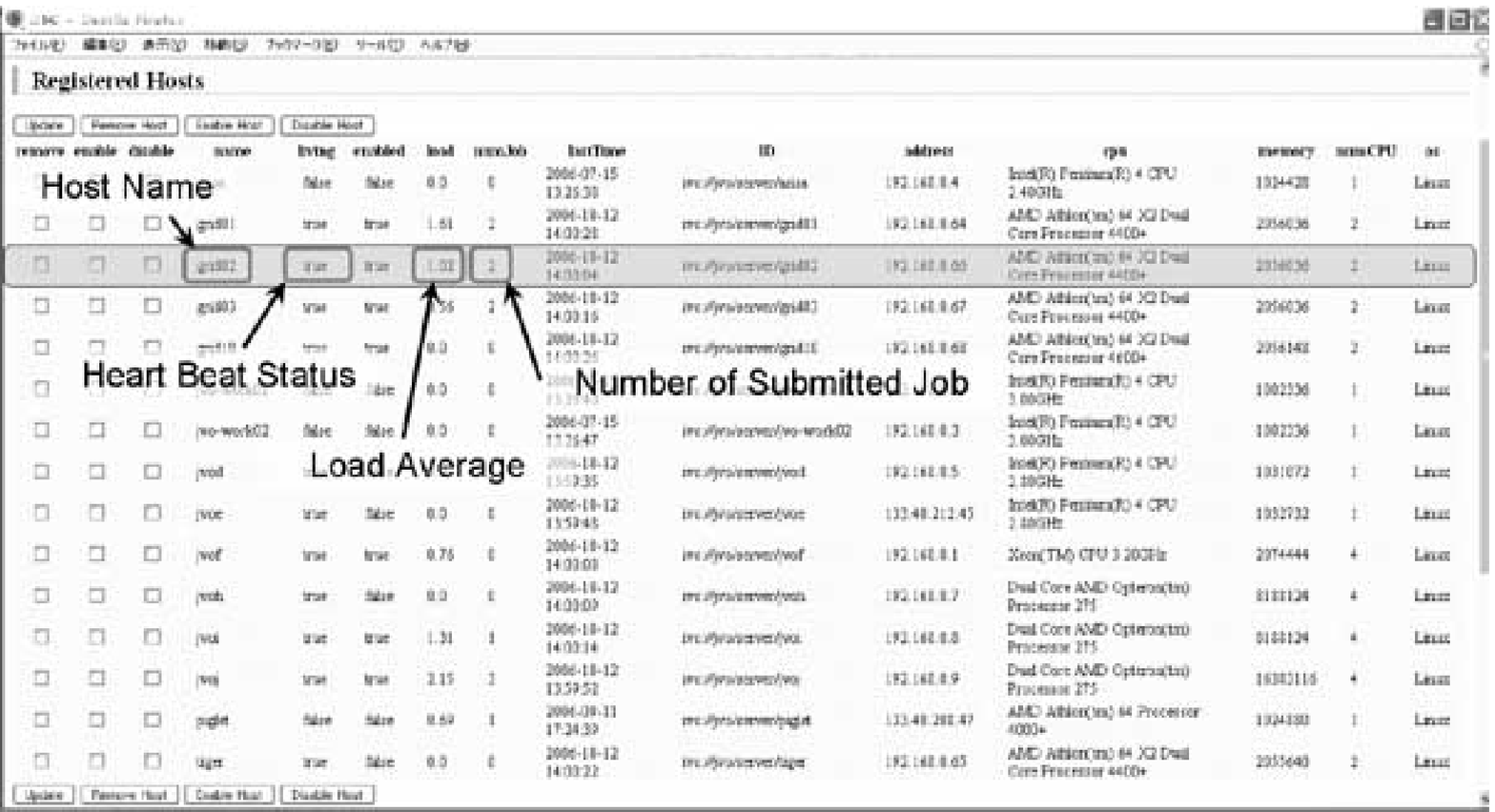}
\caption{MDS GUI} 
\label{fig:MDS}
\end{figure}
The {\bf Observation} section of the GUI is used for browsing the number 
of exposures for each filter on each day.
The observation period for calculating a response frame is determined here.
Usually select a period of successive SuprimeCam observations, which
typically lasts one week.
At the section of {\bf Create Flat Calibration Frames}, one can select
a filter and/or a chip for which a flat calibration frame is calculated.
After submitting the jobs, the job status can be viewed at the section of
{\bf Job Status}.
The status changes from ``running'' to ``data transfer'' and then 
to ``finished'' if an error does not happen.

The MDS server manages a database that contains static information (host
name, IP address, CPU type, memory size) and status (load average,
number of submitted jobs, living and enabled flags) of each DAS server, 
which can been seen on the MDS web page as shown in
Figure~\ref{fig:MDS}.
The living flag stays ``true'' while the MDS are receiving a hear beat
message from the DAS server.
When a successive five minutes of disconnection happens, the flag
changes to ``false'', and the server is recognized  as ``unavailable''.
The ``enabled'' flag is used to prohibit the job submission to the
server.
When the flag is ``false'', a job is not submitted to the server.

\section{Japanese Virtual Observatory}

\begin{figure}
\epsscale{0.8}
\plotone{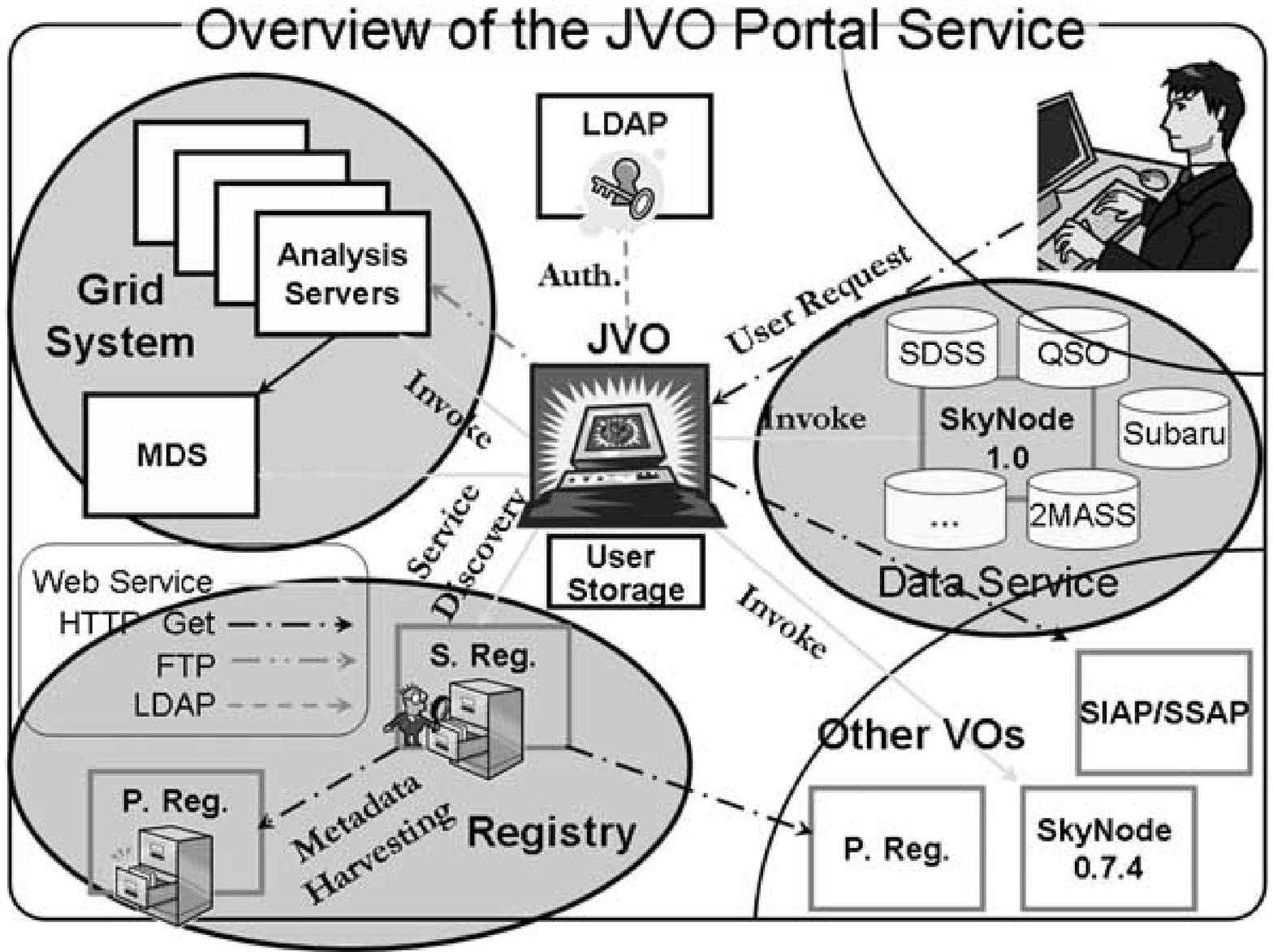}
\caption{Overview of JVO} 
\label{fig:overviewJVO}
%\end{figure}
\vspace{1em}
%\begin{figure}
\epsscale{1.0}
\plotone{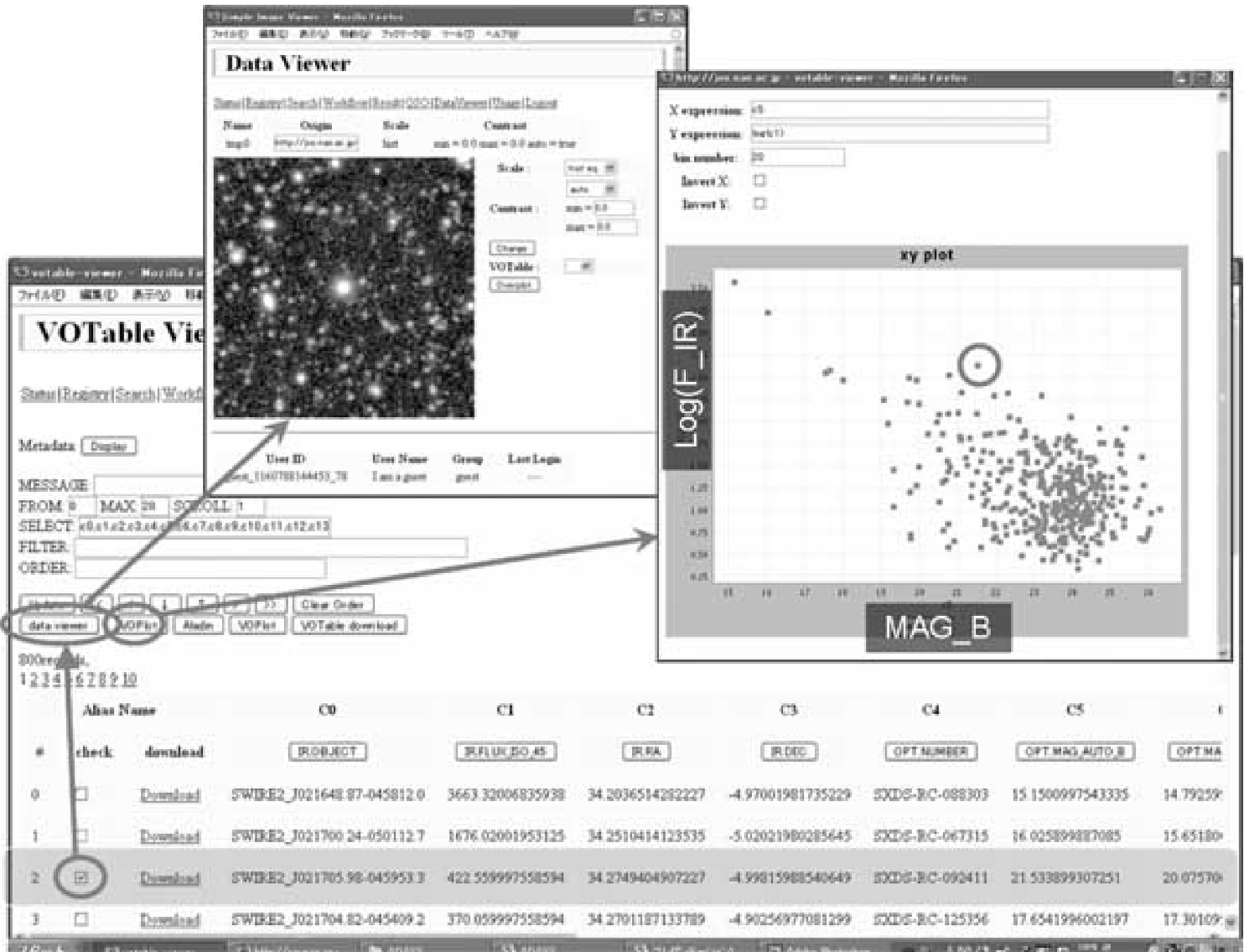}
\caption{Search result of Subaru and Swire cross match query} 
\label{fig:searchResult}
\end{figure}

The Japanese Virtual Observatory
(\htmladdnormallinkfoot{JVO}{http://jvo.nao.ac.jp/portal})
is a VO portal service, which federates the distributed VO
services and provides data analysis environment through a
web browser.
JVO is especially going to provide analysis environment utilizing 
the Subaru data.
The Subaru data currently available on the JVO are:
Subaru Deep Field catalog (SDS, SXDS), 
SuprimeCam chip image,
and SuprimeCam mosaic image.
We are preparing to provide reduced data of the other Subaru instruments.
We are also going to implement the data analysis interface for:
reduction of raw data with user specified parameters,
image manipulation,
spectrum fitting,
catalog creation from a image,
and so on.
These data analysis will be executed on the server side, so any
plugin tools are required to be installed on the user's machine.
We are going to integrate the GRID computing system described above
to the JVO portal so that enough amount of computing resource is
provided to a user.
An overview of the JVO system is shown in Figure~\ref{fig:overviewJVO}.
The details of the JVO system were described in elsewhere 
(Shirasaki et al. 2006, Shirasaki et al. 2006b, Tanaka et al. 2006, 
Ohishi et al. 2006).

Figure~\ref{fig:searchResult} shows the result of cross match query
between the catalogs of Subaru SuprimeCam (optical) and Spitzer
(infrared). 
The JVO provides several visualization tools such as JVO Data Viewer, 
JVOPlot, 
\htmladdnormallinkfoot{VOPlot}{http://vo.iucaa.ernet.in/~voi/voplot.htm}
by VO India and 
\htmladdnormallinkfoot{Aladin}{http://aladin.u-strasbg.fr/}
by CDS.
The VOPlot and Aladin are Java applets.
The JVO Data Viewer and JVOPlot are server side web applications, 
so the data itself are not transfered to the local machine; they
are converted to compact-sized graphics and transfered to the user's
web browser.
It is especially convenient to view the large data set.

\begin{figure}
\epsscale{1.0}
\plotone{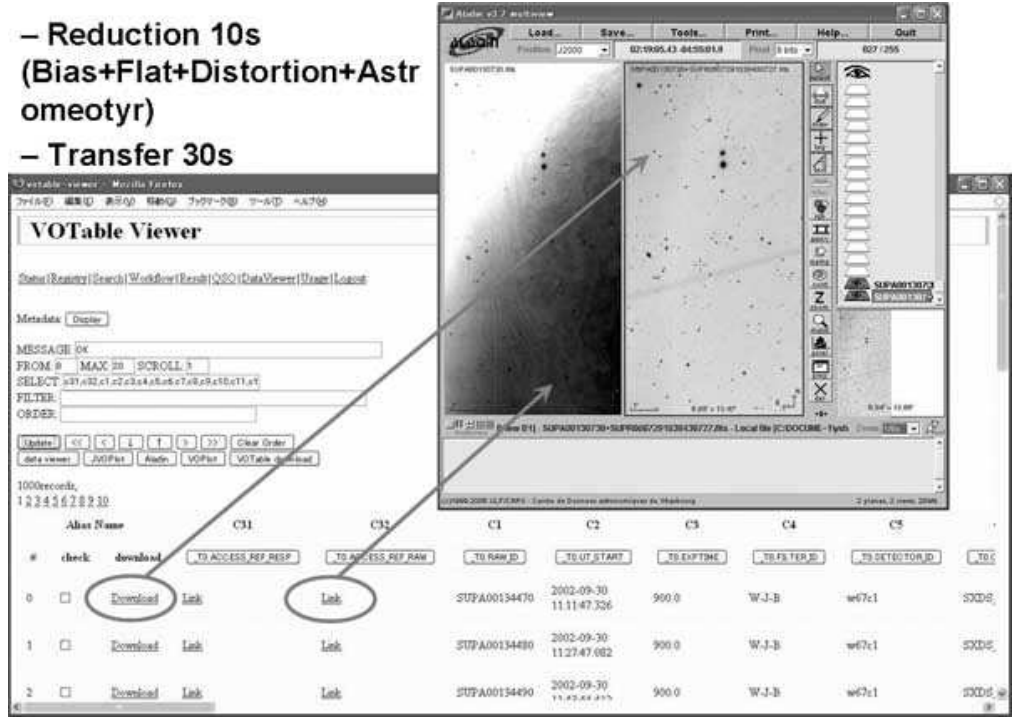}
\caption{Search result of the reduced SuprmeCam ccd frame.} 
\label{fig:subaruDataService}
\end{figure}

Figure~\ref{fig:subaruDataService} shows the result of query for
reduced SuprimeCam data.
The service returns a link to the reduced data, the link invokes 
the on-the-fly data reduction web application, and then a FITS file the
reduced image is returned.
The search result includes also the links to the raw and response frames.
The reduction procedure consists of bias subtraction, flat fielding,
distortion correction, and astrometric calibration.
The flat frames are calculated in advance by the response calculation
service described above, and registered on the database.
The typical time to execute the reduction is about 10 sec.
Once the reduced data are generated, it is stored on a cash area to
avoid repetitions of the same reduction on the same data.

\section{Summary}

The construction of the Subaru advanced data and analysis service has
just started this year (2006).
Currently the data of SuprimeCam is a primary target of the development,
but data of other instruments will be available from the JVO.
The grid computing system are constructed to obtain enough computing
resource to analyze all the SuprimeCam data in reasonably short time.
The JVO portal service is available for everyone at
\verb|http://jvo.nao.ac.jp/portal|, where currently only the limited
functionality is publicly available.
To use all the JVO functionality, user registration will be required.

\acknowledgments

This work was supported by the JSPS Core-to-Core Program,
Grant-in-aid for Information Science (15017289 and 18049074)
and Young Scientists (B) (17700085) carried out by the Ministry of
Education, Culture, Sports, Science and Technology of Japan.

\end{document}